\documentclass[aps,pre,amsfonts,amssymb,epsfig,twocolumn,groupedaddress,showpacs]{revtex4}
\usepackage{graphics}
\usepackage{epsfig}
\usepackage{amsmath}

\begin{document}

\newcommand{\beq}{\begin{eqnarray}}
\newcommand{\eeq}{\end{eqnarray}}

\title{Nonadditive entropy reconciles the area law in quantum
systems \\ with classical thermodynamics}

\author{Filippo Caruso$^{1} $ and Constantino Tsallis$^{2,3} $}

\affiliation{$^{1}$ NEST CNR-INFM and Scuola Normale Superiore,
Piazza dei Cavalieri 7, I-56126 Pisa, Italy\\
$^{2}$ Centro Brasileiro de Pesquisas Fisicas, Rua Xavier Sigaud
150, 22290-180 Rio de Janeiro, Brazil\\
$^{3}$ Santa Fe Institute, 1399 Hyde Park Road, Santa Fe, New
Mexico 87501, USA}

\begin{abstract}
The Boltzmann-Gibbs-von Neumann entropy of a large part (of linear
size $L$) of some (much larger) $d$-dimensional quantum systems
follows the so-called {\it area law} (as for black holes), i.e.,
it is proportional to $L^{d-1}$. Here we show, for $d=1,2$, that
the (nonadditive) entropy $S_q$ satisfies, for a special value of
$q \neq 1$, the classical thermodynamical prescription for the
entropy to be {\it extensive}, i.e., $S_q \propto L^d$. Therefore,
we reconcile with classical thermodynamics the area law widespread
in quantum systems. Recently, a similar behavior was exhibited, by
M. Gell-Mann, Y. Sato and one of us (C.T.), in mathematical models
with scale-invariant correlations. Finally, we find that the
system critical features are marked by a maximum of the special
entropic index $q$.
\end{abstract}


\maketitle

The aim of statistical mechanics is to establish a direct link
between mechanical microscopic laws and classical thermodynamics.
The most famous classical theory in this field was developed by
Boltzmann and Gibbs (BG) and it is considered as one of the
cornerstones of contemporary physics. The connection between the
micro- and macro-world is usually described by the so-called BG
entropy,
 \begin{eqnarray}
S_{BG} = - k \sum_{i=1}^{W} \,  p_i \ln p_i \;,
 \end{eqnarray}
where $k$ is a positive constant, $W$ is the number of microscopic
states, and $\{ p_i \}_{i=1, \ldots , W}$ is a normalized
probability distribution. The subtle concept of entropy lays the
foundation of classical thermodynamics. The BG entropy is
\textit{additive}, i.e. $S_{BG}(A,B)=S_{BG}(A)+S_{BG}(B)$, where
$A$ and $B$ are two probabilistically independent subsystems. One
of the crucial properties of the entropy in the context of
classical thermodynamics is {\it extensivity}, namely
proportionality with the number of elements of the system, when
this number is large. The BG entropy satisfies this prescription
{\it if} the subsystems are statistically (quasi-)independent, or
typically if the correlations within the system are generically
local. In such cases the system is called \textit{extensive}.

In general, however, the situation is {\it not} always of this
type and correlations may be far from negligible at all scales. In
such cases the BG entropy (of the entire system or large part of
it) may be nonextensive. Nonetheless, for an important class of
such systems, an entropy exists that is extensive in terms of the
microscopic probabilities~\cite{tsallis05}. The additive BG
entropy can be generalized into the \textit{nonadditive}
$q$-entropy~\cite{tsallis88,tsallis08}
 \begin{eqnarray}
S_q = k \frac{1-\sum_{i=1}^W p_i^q}{q-1} \, , \quad q \in {\cal
R}\;\;\;\;\;\;(S_1=S_{BG}) \; .
 \end{eqnarray}
This is the basis of the so-called {\it nonextensive statistical
mechanics}~\cite{TG,review05} (see \cite{biblio} for a regularly
updated bibliography), which generalizes the BG theory. In the
context of cybernetics and information theory, the same type of
entropic form has been advanced in \cite{others}.

Additivity (for two probabilistically independent subsystems $A$
and $B$) is generalized by the following {\it pseudo-additivity}:
$S_q(A,B)/k = S_q(A)/k + S_q(B)/k + (1-q) S_q(A) S_q(B) / k^2$.
For subsystems that have special probability correlations,
extensivity is not valid for $S_{BG}$, but may occur for $S_q$
with a particular value of the index $q \ne 1$, called the
$q$-entropic index $q_{ent}$. Such systems are sometimes referred
to as \textit{nonextensive} \cite{tsallis05,review05}. The
physical relevance of the $q$-entropy is related also to the
recent generalization of the central limit theorem (q-CLT), where
$q$-Gaussian distributions take the place of the usual Gaussians
when considering systems with strong correlations \cite{qclt}.
Much experimental evidence of predictions of nonextensive
statistical mechanics is available: see, for instance,
\cite{renzoni,goree}. Let us emphasize the difference between {\it
additivity} and {\it extensivity} for the entropy. Additivity
depends only on the mathematical definition of the entropy;
therefore, $S_1$ is additive, whereas $S_q$ ($q \ne 1$) is
nonadditive. Extensivity is more subtle, since it also depends on
the specific system, as we will show in this paper.

A physical system may exhibit genuine quantum aspects. In
particular, quantum correlations, quantified by entanglement, can
be present. The classical probability concepts are replaced by the
density matrix operator $\hat{\rho}$, in a more general
probability amplitude context. The quantum counterpart of the BG
entropy, which is called the von Neumann entropy, is thus given by
$S_1 (\hat{\rho}) = - k {\rm Tr} \, \hat{\rho} \ln \hat{\rho} $,
while the classical $q$-entropy is replaced by:
 \beq S_q (\hat{\rho}) =
k \frac{1-{\rm Tr} \, \hat{\rho}^q}{q-1} \, . \label{eq:quant_ent}
 \eeq
The pseudo-additivity is now given by
 \begin{eqnarray}
\frac{S_q (\hat{\rho}_1 \otimes \hat{\rho}_2)}{k} = \frac{S_q
(\hat{\rho}_1)}{k} + \frac{S_q (\hat{\rho}_2)}{k} + (1-q)
\frac{S_q (\hat{\rho}_1)}{k} \frac{S_q (\hat{\rho}_2)}{k} \; ,
 \nonumber\end{eqnarray}
from now on k=1.

In the following we will analyze fermionic and bosonic quantum
systems in which strong non--classical correlations exist between
their components. The appearance of long-range correlations in the
ground state of a quantum many-body system, undergoing, for
instance, a quantum phase transition at zero temperature, is
deeply related to entanglement \cite{sachdev00}. Quantum spin
chains, composed of a set of localized spins coupled through
exchange interaction in an external transverse magnetic field,
capture the essence of these intriguing phenomena and have been
extensively studied
\cite{osterloh02,osborne02,vidal03,its06,latorre04,amico}. The
degree of entanglement between a block of $L$ contiguous spins and
the rest of the chain in its ground state is measured by the von
Neumann entropy of the block. For large block size, it typically
saturates off criticality, whereas it is logarithmically unbounded
at the critical point.

Here we show that the {\it nonadditive} entropy
\cite{tsallis88,review05} $S_q(\hat{\rho}_L) \equiv k \frac{1-{\rm
Tr} \hat{\rho}_L^q}{q-1}$, $\hat{\rho}_L$ being the density matrix
of a block of $L$ spins of the ground state of quantum spin chains
in the neighborhood of a quantum phase transition, is {\it
extensive} [\textit{i.e.}, $S_q(\hat{\rho}_L) \propto L$ for $L
\gg 1$] for special values of $q<1$. The {\it additive} von
Neumann entropy $S_1(\hat{\rho}_L)=-k \,{\rm Tr} \hat{\rho}_L \ln
\hat{\rho}_L$ is {\it nonextensive} [indeed, $\lim_{L \to \infty}
S_1(\hat{\rho}_L)/L = 0$] in all considered cases; the same
happens with the {\it additive} Renyi entropy.

A similar behavior can be observed for another important class of
quantum systems, harmonic lattice Hamiltonians, \textit{i.e.},
bosons interacting through a quadratic Hamiltonian in dimension
$d=2$. These systems are discrete versions of the free scalar
Klein-Gordon field. The degree of entanglement between a square
block of $L^2$ oscillators and its exterior, as measured by the
\textit{additive} von Neumann entropy of the square block,
$S_1(\hat{\rho}_{L^2})$ is linear in $L$ (\textit{area law}), i.e.
it is \textit{nonextensive}
\cite{barthel,plenio,bombelli,Srednicki,tHooft}. Here we show that
the \textit{nonadditive} entropy $S_q(\hat{\rho}_{L^2})$ is
\textit{extensive} [\textit{i.e.}, $S_q(\hat{\rho}_{L^2}) \propto
L^2$ for $L \gg 1$] for special values of $q<1$.

Therefore, we present here two physical realizations, in many-body
Hamiltonian systems, of the abstract mathematical examples
recently exhibited by M. Gell-Mann, Y. Sato and one of us (C.T.)
in Ref. \cite{tsallis05}, that, for anomalous values of $q$, the
nonadditive entropy $S_q$, can be extensive, as expected from the
Clausius thermodynamical requirement for the entropy. In other
terms, we show explicitly the first two physical examples of the
fact that the nonadditive entropy reconciles the area law (e.g.,
typical of black holes) with classical thermodynamics. In addition
to the considerable advantage of $S_q$ enabling thus the use of
all standard thermodynamical relations, this constitutes a
powerful tool to detect strong nonlocal correlations in quantum
many-body systems, by a nonadditive measure
\cite{zyczkoski,zeilinger,grigolini,lloyd,abe,rajagopal,virmani}.
Indeed, this entropic index $q$ presents a maximum when the
correlation length is divergent in the ground state of these
quantum systems.

\section{Quantum spin chains}

First of all, we focus our investigations on a one-dimensional
spin-1/2 ferromagnetic chain with an exchange (local) coupling and
subjected to an external transverse magnetic field, \textit{i.e.},
the quantum XY model. The Hamiltonian is given by:
 \beq \hat{\mathcal H} =
- \sum_{j=1}^{N-1} \left[ (1 + \gamma) \hat{\sigma}^x_j
\hat{\sigma}^x_{j+1} + (1 - \gamma) \hat{\sigma}^y_j
\hat{\sigma}^y_{j+1} + 2 \lambda \hat{\sigma}^z_j \right]
\label{eq:XYham} \eeq where $\hat{\sigma}^\alpha_j (\alpha =
x,y,z)$ are the Pauli matrices of the $j$-th spin, $N$ is the
number of spins of the chain, and $\gamma$ and $\lambda$
characterize, respectively, the strength of the anisotropy
parameter and of a trasverse magnetic field along the $z$
direction. This model for $0 < \vert \gamma \vert \leq 1$ belongs
to the Ising universality class and it actually reduces to the
quantum Ising chain for $|\gamma|=1$. At $T=0$, this system
undergoes a quantum phase transition at the critical point
$|\lambda_c| = 1$ in the thermodynamic limit $N \to \infty$. For
$\gamma=0$ it is the isotropic XX model, which is critical for
$\vert \lambda \vert \leq 1$~\cite{sachdev00}.

Entanglement in the neighborhood of the quantum phase transition
has been recently widely investigated
\cite{osterloh02,osborne02,vidal03,its06,latorre04,amico}. In
particular it has been shown that one-site and two-site
entanglement between nearest or next-to-nearest spins display a
peak near or at the critical point~\cite{osterloh02,osborne02}.
Moreover, the entanglement between a block of $L$ contiguous spins
and the rest of the chain in the ground state, quantified by the
von Neumann entropy, presents a logarithmic divergence with $L$ at
criticality, while it saturates in a non--critical
regime~\cite{vidal03,its06,latorre04}.

The inadequacy of the additive von Neumann entropy as a measure of
the information content in a quantum state was pointed out in Ref.
\cite{zeilinger}. A theoretical observation that the measure of
quantum entanglement may not be additive has been discussed in
Refs.
\cite{grigolini,zyczkoski,zeilinger,lloyd,abe,rajagopal,virmani}.
Recently, Ref. \cite{popescu} suggested to abandon the \textit{a
priori} probability postulate going beyond the usual BG situation.

Here we propose to extend the definition of the von Neumann
entropy to a wider class of entropy measures (see also Ref.
\cite{zyczkoski2}) which naturally include it, thus generalizing
the notion of the block entanglement entropy. The block
$q$-entropy of a block of size $L$ is simply defined as the
$q$-entropy, Eq.~(\ref{eq:quant_ent}), of the reduced density
matrix $\hat{\rho}_L$ of the block, $S_q(\hat{\rho}_L)$, when the
total chain is in the ground state. In the following we show that,
in contrast to the von Neumann entropy, there exists a $q$ value
for which $S_q(\hat{\rho}_L)$ is {\it extensive}. This value does
depend on the critical properties of the chain and it is
consistent with the universality hypothesis.

The XY model in Eq.~(\ref{eq:XYham}) can be diagonalized exactly
with a Jordan-Wigner transformation, followed by a Bogoliubov
rotation~\cite{lieb61,pfeuty,barouch,latorre04}; this allows one
to analytically evaluate the spectrum of $\hat{\rho}_L$ in the
thermodynamic limit $N \to \infty$. More details are shown also in
Ref. \cite{carusoct}.

\begin{figure}
\begin{center}
\includegraphics[width=0.45\textwidth]{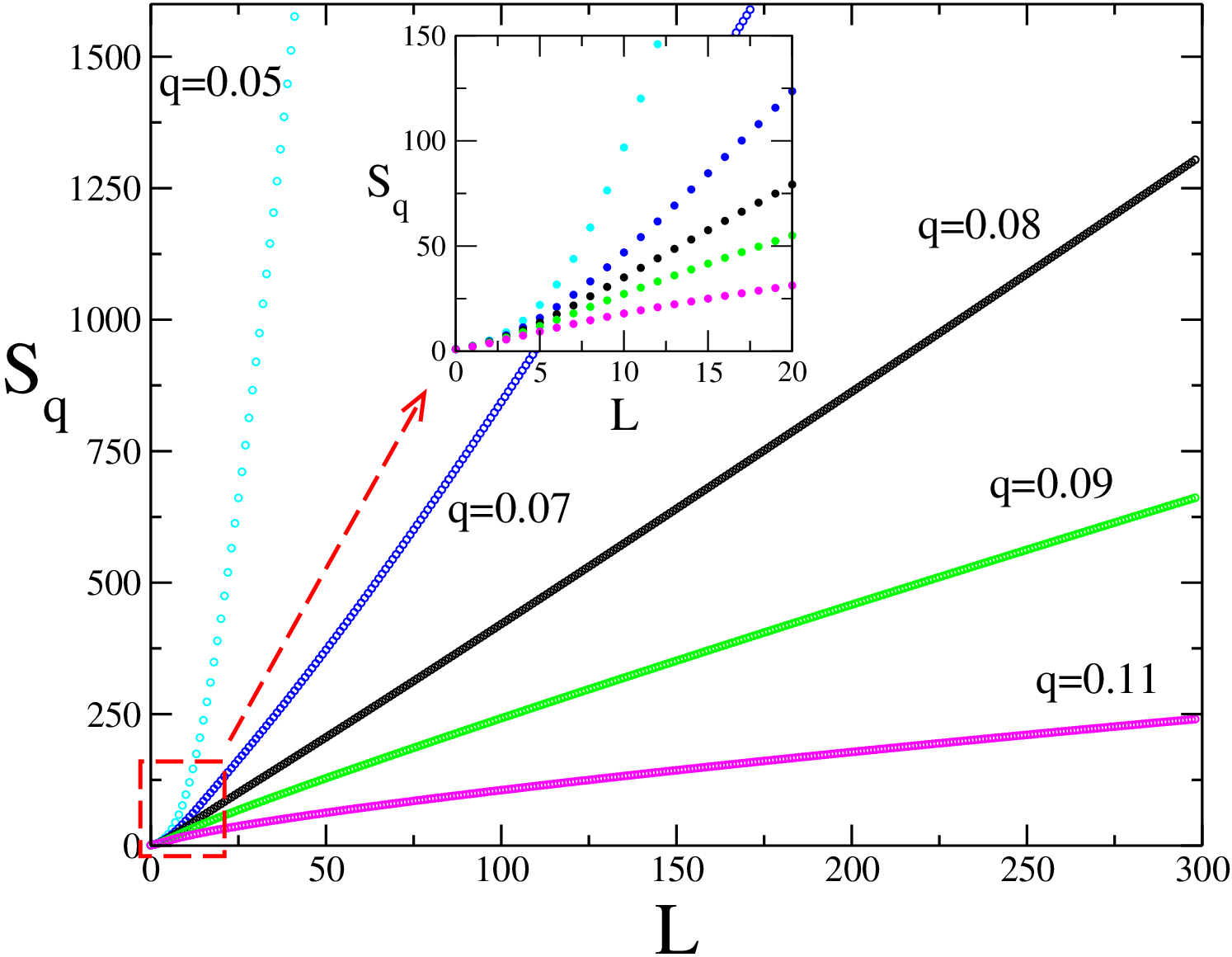} \caption{Block $q$-entropy
$S_q(\hat{\rho}_L)$ as a function of the block size $L$ in a
critical Ising chain ($\gamma=1, \, \lambda=1$), for typical
values of $q$. Only for $q=q_{ent} \simeq 0.0828$ is $s_q$ {\it
finite} (\textit{i.e.}, $S_q$ is {\it extensive}); for $q <
q_{ent}$ ($q > q_{ent}$) it diverges (vanishes).}\label{fig.1}
\end{center}
\end{figure}

We first analyze the anisotropic quantum XY model,
Eq.~(\ref{eq:XYham}) with $\gamma \neq 0$, which has a critical
point at $\lambda_c = 1$. The block $q$-entropy as a function of
the block size can show completely different asymptotic behaviors,
when the entropic index $q$ is varied. In particular, here we are
interested in a thermodynamically relevant quantity, namely the
slope, denoted $s_q$, of $S_q$ versus $L$. It is generically not
possible to have a finite value of $s_1$: the entanglement
entropy, evaluated by the von Neumann entropy, either saturates or
diverges logarithmically in the thermodynamic limit, for
respectively non--critical or critical spin chains
\cite{vidal03,its06,latorre04}. The situation dramatically changes
when the $q$-entropy formula in Eq.~(\ref{eq:quant_ent}) is used:
qualitatively it happens that, regardless the presence or absence
of criticality, a $\lambda$-dependent value of $q$, denoted
$q_{ent}$, exists such that, in the range $1 \ll L \ll \xi$ ($\xi$
being the correlation length), $s_{q_{ent}}$ is finite, whereas it
vanishes (diverges) for $q>q_{ent}$ ($q<q_{ent}$). We note that
here the nonextensivity (\textit{i.e.}, $q\neq 1$) features are
not due to the presence of say long-range interactions
\cite{anteneodo,rpt} but they are triggered only by the fully
quantum {\it nonlocal} correlations. In Fig.~\ref{fig.1} we show,
for the critical Ising model ($\gamma=1$, $\lambda=1$), the
behavior of the block $q$-entropy with respect to the block size:
$S_q(\hat{\rho}_L)$ becomes extensive [\textit{i.e.}, $0< \lim_{L
\to \infty} S_q(\hat \rho_L)/L < \infty$] at a $q$-entropic index
$q_{ent} \simeq 0.0828 \pm 10^{-4}$ (with a corresponding entropic
density $s_{q_{ent}} \approx 3.56 \pm 0.03$), thus satisfying the
prescriptions of classical thermodynamics.

\begin{figure}
\begin{center}
\includegraphics[width=0.45\textwidth]{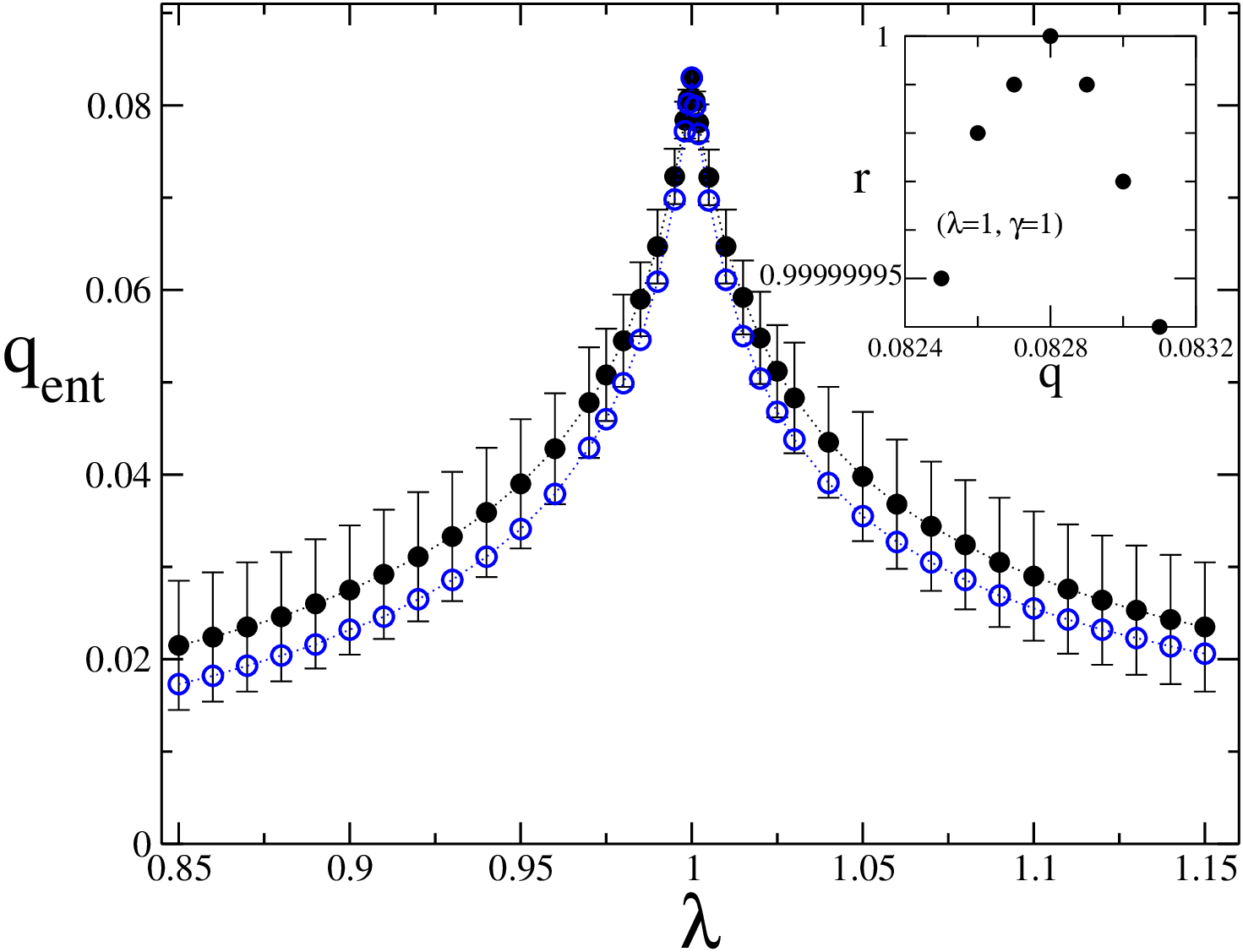} \caption{The
$\lambda$-dependence of the index $q_{ent}$ in the Ising
($\gamma=1$, solid circle) and XY ($\gamma=0.75$, empty circle)
chains. {\it Inset}: determination of $q_{ent}$ through numerical
maximization of the linear correlation coefficient $r$ of
$S_q(\hat{\rho}_L)$. The error bars for the Ising chain are
obtained considering the variation of $q_{ent}$ when using the
range $100 \le L \le 400$ in the search for $S_q(\hat{\rho}_L)$
linear behavior. At the present numerical level, a vanishing $q$
cannot be excluded off-criticality because of finite-size
effects.}\label{fig.2a}
\end{center}
\end{figure}

\begin{figure}[h!]
\includegraphics[width=0.43\textwidth]{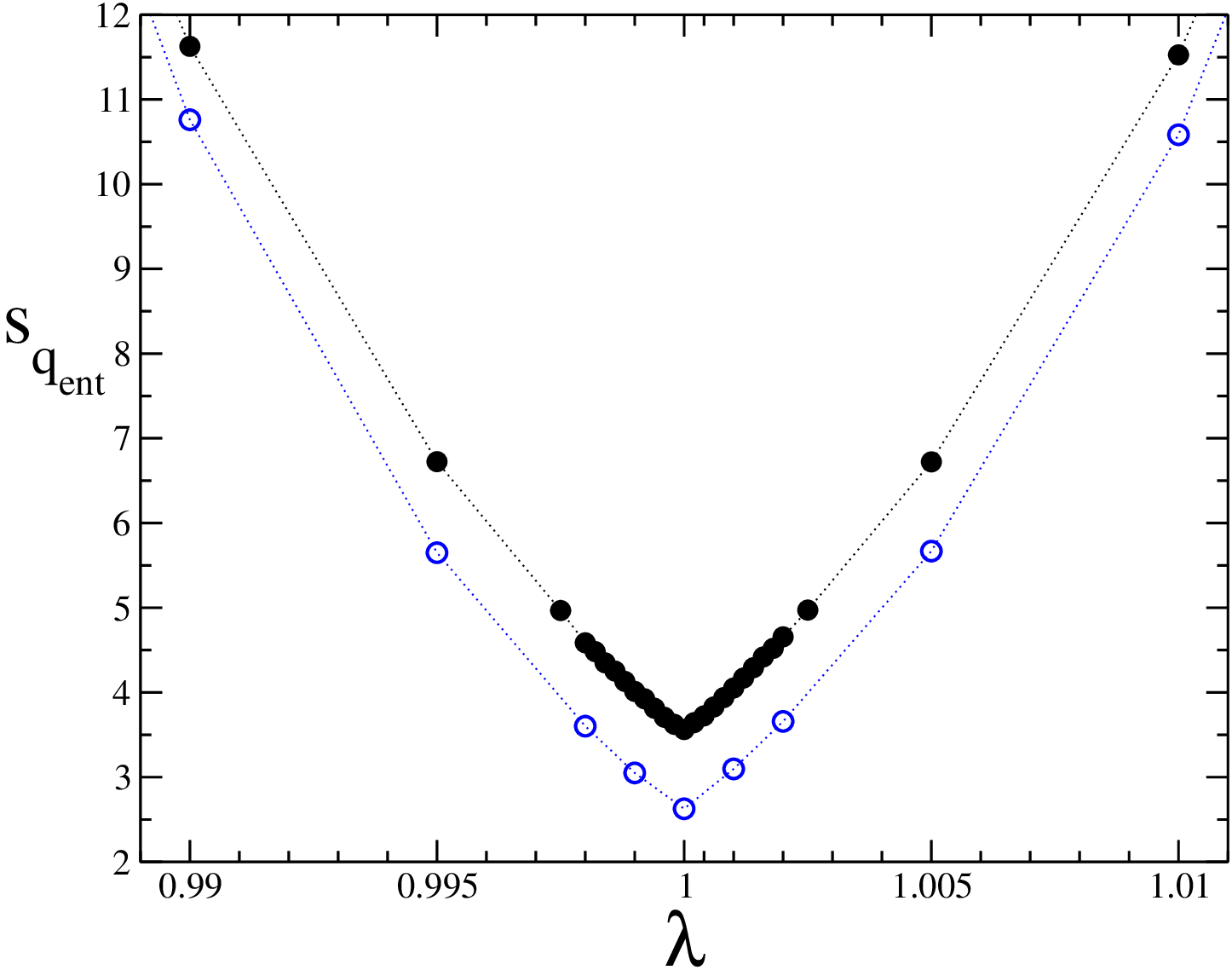}
\caption{$\lambda$-dependence of the $q$-entropic density $s_{q_{ent}}$ in
the Ising ($\gamma = 1$, solid circle) and XY ($\gamma = 0.75$,
empty circle) models. For $\lambda=1$, the slopes are $3.56$ and
$2.63$, for $\gamma =1$ and $\gamma=0.75$, respectively. Notice
the considerable variation in the values of the slope for even
slightly different values of $\lambda$ out of criticality.
}\label{fig.2b}
\end{figure}

A very similar behavior is shown for the non--critical Ising
model, as well as for critical and non--critical XY models with
$0<\gamma<1$. The value of $q_{ent}$, for which
$S_q(\hat{\rho}_L)$ is asymptotically extensive, is obtained by
maximizing numerically the linear correlation coefficient $r$ of
$S_q(\hat{\rho}_L)$, in the range $1 \ll L \ll \xi$, with respect
to $q$, as shown in the inset in Fig.~\ref{fig.2a}. Let us stress
that, at precisely the critical point, $\xi$ diverges, hence $L$
is unrestricted and can run up to infinity. The index $q_{ent}$
depends on the distance from criticality and it increases as
$\lambda$ approaches $\lambda_c$ (Fig.~\ref{fig.2a}). It is worth
stressing that our numerical results satisfy the duality symmetry
$\lambda \longrightarrow 1/\lambda$, investigated in Ref.
\cite{savit}. For the sake of clarity, notice that the value of
$q_{ent}$ may be vanishing off criticality and this could not be
found numerically because of the presence of finite-size effects.
We have also checked other values of $\gamma$ for the XY model and
the results are very similar to those presented here. This fact is
consistent with the universality hypothesis. On one hand, the XY
and Ising models (Ising universality class) have the same behavior
as regards the extensivity of $S_q(\hat{\rho}_L)$. In Fig.
\ref{fig.2b} we report the variation of $s_{q_{ent}}$ with respect
to $\lambda$. On the other hand, for the isotropic XX model
($\gamma=0$) in the critical region $\vert \lambda \vert \leq 1$
we find $q_{ent}\simeq 0.15 \pm 0.01$ for which $S_{q}(\hat
\rho_L)$ becomes {\it extensive}.

\begin{figure}
\begin{center}
\includegraphics[width=0.45\textwidth]{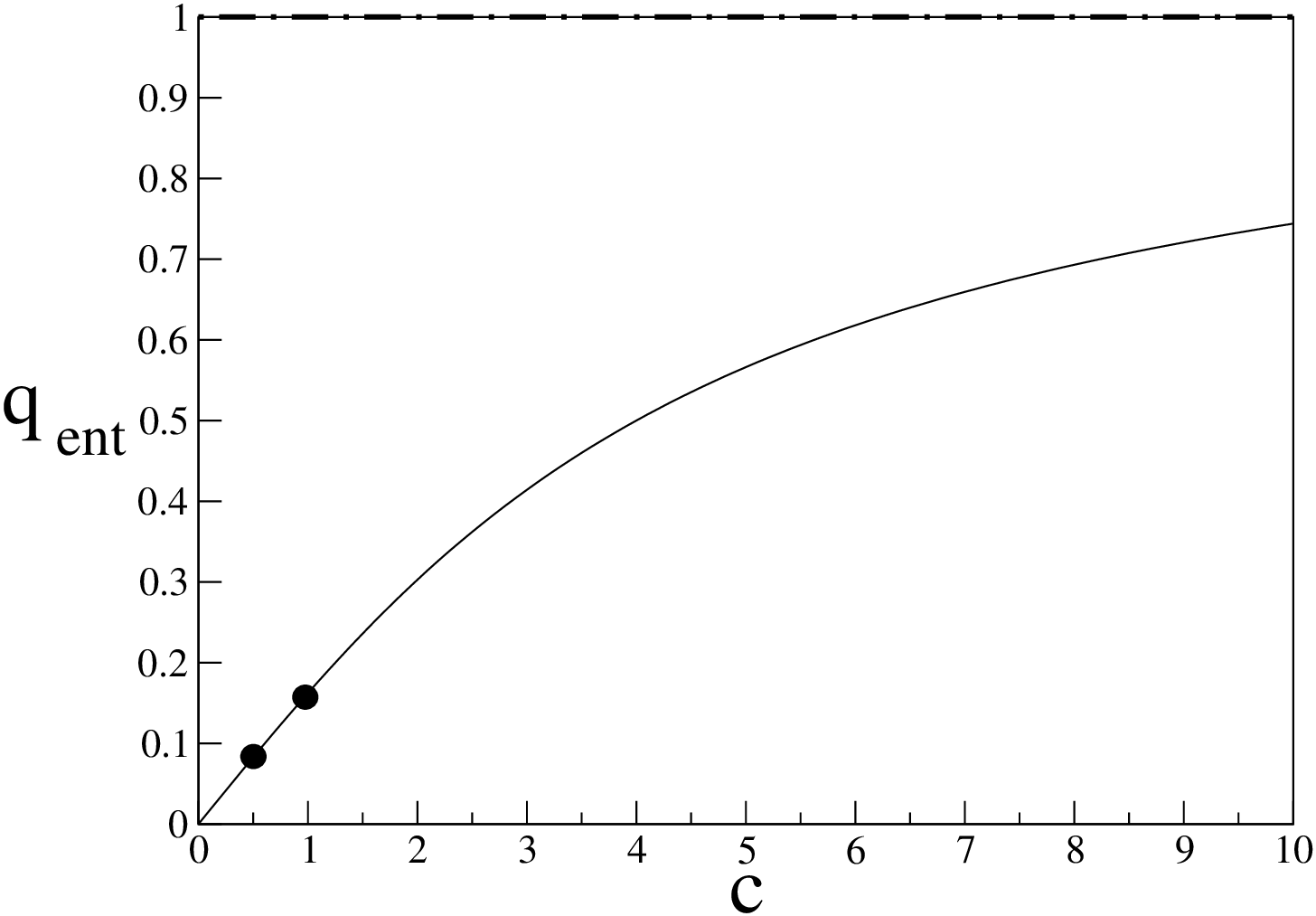}
\caption{$q_{ent}$ versus $c$ with the $q$-entropy,
$S_q(\hat{\rho}_L)$, being \textit{extensive}, \textit{i.e.},
$\lim_{L \to\infty}S_{\frac{\sqrt{9+c^2}-3}{c}}(\hat{\rho}_L)/L<
\infty$. When $c$ increases from $0$ to infinity, $q_{ent}$
increases from $0$ to unity (von Neumann entropy). For the
critical quantum Ising and XY models $c=1/2$ and $q_{ent} =
\sqrt{37} -6 \simeq 0.0828$, while for the critical isotropic XX
model $c=1$ and $q_{ent}=\sqrt{10} - 3 \simeq 0.16$.
}\label{fig.3}
\end{center}
\end{figure}

Reference \cite{cardy} enables us to analytically confirm (only at
the critical point) our numerical results. The continuum limit of
a (1+1)-dimensional critical system is a conformal field theory
with central charge $c$. In this quite different context, the
authors re-derive the result $S_1(\hat{\rho}_L)\sim (c/3) \ln L$
for a finite block of length $L$ in an infinite critical system.
To obtain the von Neumann entropy, they find an analytical
expression for ${\rm Tr} \hat{\rho}^q_L$, namely $ {\rm Tr}
\hat{\rho}_L^q \sim L^{-c/6(q-1/q)}$. Here, we use this expression
quite differently. We impose the extensivity of
$S_q(\hat{\rho}_L)$ and we find the value of $q$ for which
$-c/6(q_{ent}-1/q_{ent})=1$, \textit{i.e.},
\begin{equation}q_{ent} = \frac{\sqrt{9+c^2}-3}{c} \; ; \label{charge}
\end{equation}
consequently, $\lim_{L
\to\infty}S_{\frac{\sqrt{9+c^2}-3}{c}}(\hat{\rho}_L)/L< \infty$.
When $c$ increases from $0$ to infinity (see Fig.~\ref{fig.3}),
$q_{ent}$ increases from $0$ to unity (von Neumann entropy). It is
well known that for the critical quantum Ising and XY models the
central charge is equal to $c=1/2$ (indeed they are in the same
universality class and can be mapped to a free fermionic field
theory). For these models, at $\lambda=1$, the value of $q$ for
which $S_q(\hat{\rho}_L)$ is {\it extensive} is given by $q_{ent}
= \sqrt{37} -6 \simeq 0.0828$, in perfect agreement with our
numerical results in Fig.~\ref{fig.2a}. The critical isotropic XX
model ($\gamma=0$ and $|\lambda| \leq 1$) is, instead, in another
universality class, the central charge is $c=1$ (free bosonic
field theory) and $S_q(\hat{\rho}_L)$ is {\it extensive} for
$q_{ent} = \sqrt{10} -3 \simeq 0.16$, as found also numerically.
Therefore, the universal behavior of the $q$-entropic index
$q_{ent}$ is strictly related to the universal role played by the
central charge in conformal field theory. Eq.~(\ref{charge})
represents an additional connection between nonextensive
statistical mechanical concepts and BG statistical mechanics at
criticality. See Ref. \cite{robledo} for another connection, where
once again we verify that the $q$-entropic index typically
characterizes universality classes. Let us note that, when the
critical one-dimensional (1D) system is a semi-infinite chain, one
has to replace $c$ with $c/2$ in Eq.~(\ref{charge}) \cite{cardy}.

It is worth mentioning that the Renyi entropy of a block of
critical XX spin chains has been derived analytically in Refs.
\cite{jin04,franchini}. Since the Renyi entropy is simply
connected to the $q$-entropy, it is possible to re-derive
$q_{ent}$ for the critical XX model also from that analytical
expression. Finally, let us point out that the reduction of the
{\it pure} ground state of the full chain (at $T=0$) to a finite
block of $L$ spins results in a {\it  mixed} state with quantum
fluctuations. A mapping of this subsystem within a
zero-temperature XX infinite chain to a finite system which is
thermalized at some finite temperature has been recently exhibited
\cite{eisler}, thus defining an $L$-dependent effective
temperature of the block. The use of a non-Boltzmannian
distribution (\textit{e.g.}, the one emerging within nonextensive
statistical mechanics) might enable the definition of an effective
temperature which would {\it not} depend on $L$, as physically
desirable. Indeed, this approach has been successfully implemented
for $e-e^+$ collision experiments \cite{curado}.

\section{2D bosonic systems}

Now we present a second physical realization of the extensivity of
$S_q$ in a bosonic 2D system at $T=0$. We start from a
bidimensional (square lattice) system of infinite coupled harmonic
oscillators studied in Ref. \cite{barthel}, with Hamiltonian
\begin{footnotesize}
 \beq \label{eqn:HharmLattice}
H  =  \frac{1}{2} \sum_{x,y}\,\Big[\Pi_{x,y}^2 + \omega_0^2
\,\Phi_{x,y}^2 +  (\Phi_{x,y}-\Phi_{x+1,y})^2 +
(\Phi_{x,y}-\Phi_{x,y+1})^2\Big]
    ,\nonumber
\eeq
 \end{footnotesize}
 where $\Phi_{x,y}$, $\Pi_{x,y}$ and $\omega_0$ are the
coordinate, momentum and self-frequency of the oscillator at site
$\vec{r}=(x,y)$. The system has the dispersion relation
$E(\vec{k}) = \sqrt{\omega_0^2 + 4\sin^2{k_x/2}+4\sin^2{k_y/2}}$,
\textit{i.e.}, a gap $\omega_0$ at $\vec{k}=\vec{0}$. Applying the
canonical transformation $b_{\textbf{i}} =
\sqrt{\frac{\omega}{2}}\* (\Phi_{\textbf{i}}
   + \frac{i}{\omega} \Pi_{\textbf{i}})$ with $\omega  = \sqrt{\omega_0^2 + 4}$
   and $i=1,...,L^2$,
the Hamiltonian in Eq. (\ref{eqn:HharmLattice}) is mapped to the
quadratic canonical form
 \beq
H = \sum_{ij} \left[ a_i^{\dagger} A_{ij}a_j + \frac{1}{2}
         (a_i^{\dagger} B_{ij} a_j^{\dagger} + h.c. )\right] \; ,
 \eeq
where $a_i$ are bosonic operators and $i$, $j$ range from $1$ to
$L^2$. In Ref. \cite{barthel} the authors find an asymptotic
linear behavior (\textit{i.e., area law}) of the block
entanglement entropy as a function of the linear size $L$ (they
consider square blocks of area $L^2$), for several $\omega_0$ ($
\lim_{L \to \infty} S_1(\hat \rho_L)/L^2 =0$), no matter how close
the gap energy is to zero.

\begin{figure}
\begin{center}
\includegraphics[width=0.45\textwidth]{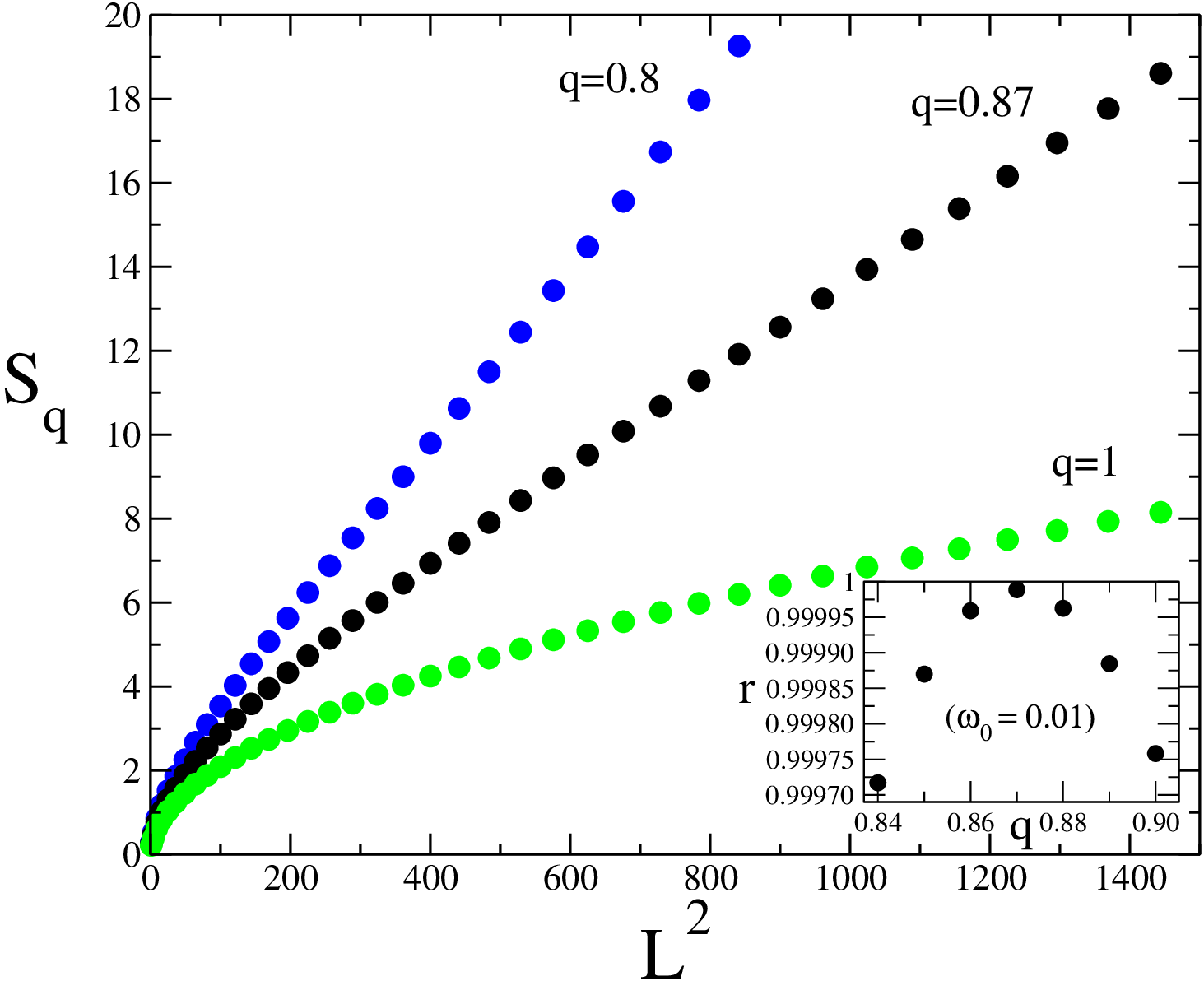}
\caption{Block $q$-entropy $S_q(\hat \rho_L)$ as a function of the
square block area $L^2$ in a bosonic 2D array of infinite coupled
harmonic oscillators at $T=0$, for typical values of $q$. Only for
$q=q_{ent} \simeq 0.87$, $s_q$ is {\it finite} (\textit{i.e.},
$S_q$ is {\it extensive}); for $q < q_{ent}$ ($q > q_{ent}$) it
diverges (vanishes). {\it Inset}: determination of $q_{ent}$
through numerical maximization of the linear correlation
coefficient $r$ of $S_q(\hat{\rho}_L)$ in the range $400 \le L^2
\le 1600$.}\label{fig.4}
\end{center}
\end{figure}

\begin{figure}
\begin{center}
\includegraphics[width=0.45\textwidth]{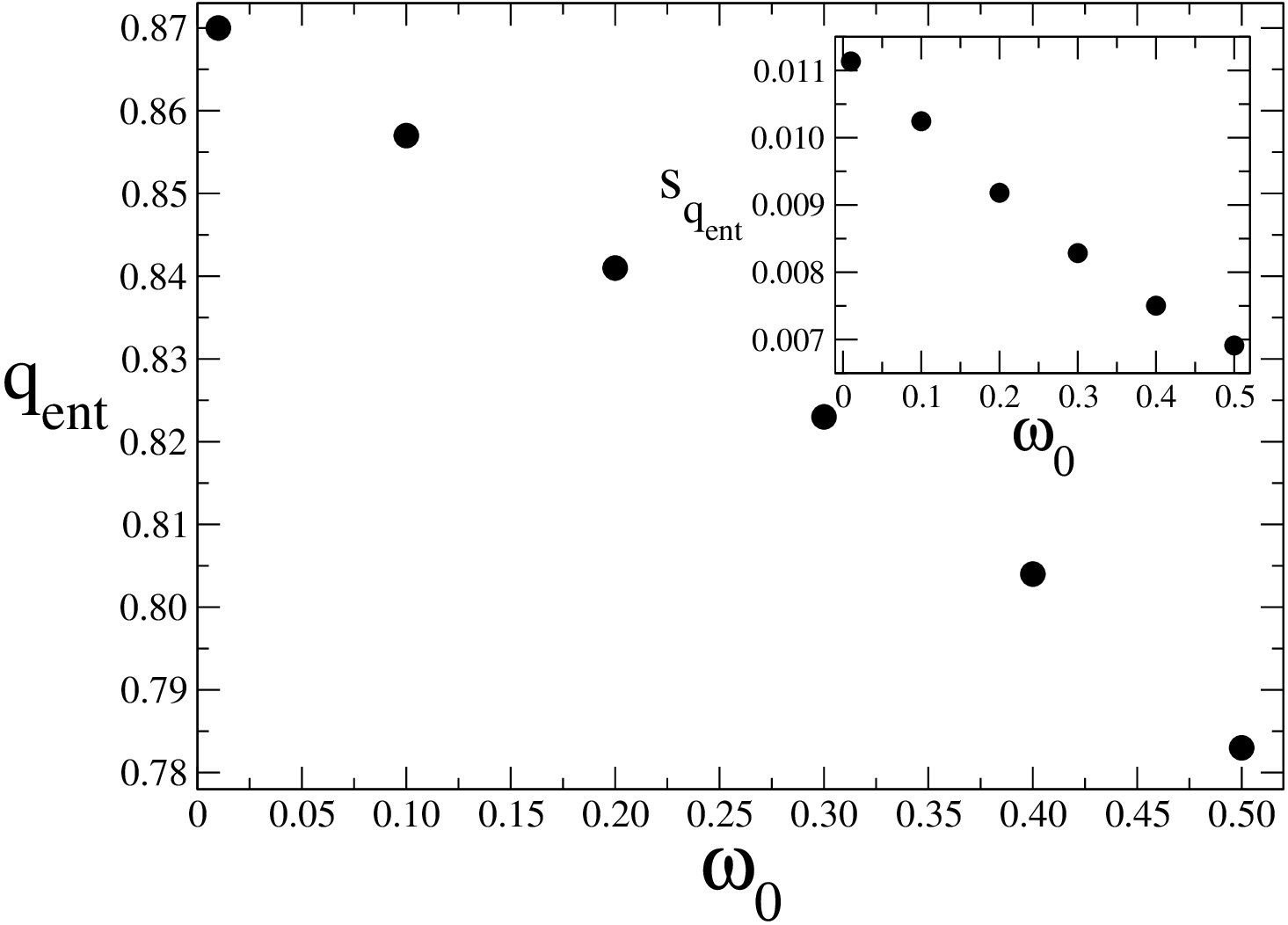}
\caption{$\omega_0$-dependence of the index $q_{ent}$ in a bosonic
2D array of infinite coupled harmonic oscillators at $T=0$. {\it
Inset:} the $\omega_0$-dependence of the $q$-entropic density
$s_{q_{ent}}$.}\label{fig.5}
\end{center}
\end{figure}

Here we study, instead, the behavior of the block $q$-entropy of
the reduced density operator of a square block as a function of
its area $L^2$, when the bosonic infinite two-dimensional system
is in its ground state. We follow a similar procedure to the one
used above for quantum spin chains (see Ref. \cite{barthel} for
more details). In Fig.~\ref{fig.4} we show, in the case of
$\omega_0=0.01$, that $S_q(\hat{\rho}_L)$ becomes {\it extensive}
(\textit{i.e.}, $0< \lim_{L \to \infty} S_q(\hat \rho_L)/L^2 <
\infty$) at an index $q_{ent} \simeq 0.87$ (with a corresponding
entropic density $s_{q_{ent}} \approx 0.011$; see inset in
Fig.~\ref{fig.5}). A very similar behavior is shown for other
values of the gap. Let us point out that, unlike the linear
behavior (for any gap energy) of the von Neumann entropy, now the
index $q_{ent}$ depends on the gap and therefore measures the
presence of a progressively divergent correlation length, as shown
in Fig.~\ref{fig.5}.

\section{Final Remarks}

We present two quantum many-body Hamiltonian physical realizations
of the mathematical probabilistic models with scale-invariant
correlations recently shown by M. Gell-Mann, Y. Sato and one of us
(C.T.) in Ref. \cite{tsallis05}, in which the nonadditive entropy
$S_q$ can be applied successfully (\textit{i.e.}, satisfying the
classical thermodynamic requirement of extensivity). In this basic
manner, we reconcile the entropy area law characterizing many
quantum systems with classical thermodynamics. In addition to
that, the present results show clearly the difference between {\it
additivity} and {\it extensivity} for the entropy. Additivity
depends only on the mathematical features of the entropy, e.g,
$S_1$ is additive while $S_q$ ($q \ne 1$) is nonadditive.
Extensivity is a more subtle concept and relies on both the
mathematical features of the entropy and the specific physical
system. Indeed, the $T=0$ block entropies of the $1/2$-spin $d=1$
quantum system at criticality are given by $S_1(L) \propto \ln L$
({\it i.e.}, nonextensive), and $S_{[\sqrt{9+c^2}-3]/c}(L) \propto
L$ ({\it i.e.}, extensive). Moreover, the $T=0$ block entropies of
the $d=2$ bosonic system are given by $S_1(L) \propto L$ ({\it
i.e.}, nonextensive), and $S_{q}(L) \propto L^2$ for special
values of $q<1$ ({\it i.e.}, extensive); for instance, $q_{ent}
\simeq 0.87$ for vanishing gap energy.

More generally, it is known (see Refs. \cite{plenio,barthel} and
references therein) that, for $d$-dimensional bosonic systems
(e.g., a black hole \cite{bombelli,Srednicki,tHooft}), $S_1$
follows the area law, {\it i.e.}, $S_1(L) \propto L^{d-1}$
(nonextensive). Let us point out that the behavior of the block
entropy $S_1(L)$ for these quantum systems matches well-known
results in conformal field theory (as noted above), where the
analogous of the block entropy is the so-called geometric entropy,
defined in the continuum \cite{bombelli,Srednicki,tHooft,cardy}.
As first suggested by 't Hooft (1985) and later shown by Callan
and Wilczek (1994), the geometric entropy is the first quantum
correction to a thermodynamical entropy, which reduces to the
Bekenstein-Hawking entropy for black holes
\cite{tHooft,bombelli,Srednicki}. A direct connection between
entropy and boundary area has been clearly suggested and
numerically implemented in Refs. \cite{bombelli,Srednicki}
(compare with Refs. \cite{plenio,barthel} and references therein).
Finally, the relation among entanglement entropy, the black hole
area law and other concepts such as the holographic bound is still
an open problem \cite{bousso}. In this context, it is interesting
to note that a logarithmic behavior for $d=1$ and the area law for
$d>1$ for a large class of fermionic and bosonic $d$-dimensional
many-body Hamiltonians with short-range interaction at $T=0$ can
be unified through $S_1(L) \propto [L^{d-1}-1]/(d-1) \equiv
\ln_{2-d} L$ ({\it i.e.}, nonextensive, $ \ln L$ for $d=1$, and $
L^{d-1}$ for $d>1$, area law) \cite{qlog}, which would correspond
to a large class (not yet completely identified) of fully
entangled quantum systems. For all these systems, one could expect
that a value of $q$ exists such that $S_q(L) \propto L^d$ ({\it
i.e.}, extensive). In this paper our conjecture is verified for
$d=1$ quantum spin chains and $d=2$ bosonic systems and is
therefore promising also in higher dimensions.

The present work has benefited from early collaboration with D.
Rossini and D. Patan\`{e}, fruitful discussions with R. Fazio, V.
Giovannetti, A. Pluchino, A. Rapisarda, and interesting related
comments by G. 't Hooft. It was partially supported by the Centro
di Ricerca Ennio De Giorgi of the Scuola Normale Superiore, and by
CNPq and Faperj (Brazilian Agencies).

\end{document}